\documentclass[journal]{IEEEtran}

\usepackage{graphicx}
\usepackage{float}
\graphicspath{{figures/}}

\usepackage{amsmath}

\usepackage{cite}
\usepackage{footnote}
\usepackage{url}
\makesavenoteenv{tabular}

\begin{document}
	\title{A PXI-based Multi-channel Data Acquisition System for Fast Transient Pulses}
	\author{Yafei~Du,~Jun~Wu,~Chen~Yuan,~Haohan~Yang,~Chuanfei~Zhang,~Yinong~Liu
		\thanks{Manuscript received June 24, 2018.}
		\thanks{Y. Du is with the Department of Engineering Physics, Tsinghua University,
			Beijing 100084, China and the Institute of Nuclear Physics and Chemistry, China Academy of Engineering Physics, Sichuan, Mianyang 621900, China. dyf13@tsinghua.org.cn.}
		\thanks{J. Wu, C. Yuan, H. Yang and C. Zhang are with the Institute of Nuclear Physics and Chemistry, China Academy of Engineering Physics,  Sichuan, Mianyang 621900, China.}
		\thanks{Y. Liu is with the Department of Engineering Physics, Tsinghua University,
			Beijing 100084, China.}}
	

	\maketitle
	\thispagestyle{empty}

	\begin{abstract}
		In this paper, we design a PXI-based, multi-channel data-acquisition system (DAS) mainly applicable to recording one-shot fast transient pulses in nuclear physics experiments.
		The system consists of one NI PXIe-1085 chassis, containing a controller card and at most 16 data-acquisition (DAQ) cards.
		Every single DAQ card has a sampling rate of 1GS/s and a 12bit vertical resolution with the PXI interface and SFP+ transceiver for data transmission.
		When the system is put into operation near the pulsed radiation source, the SFP+ optical fiber channel enables a timely data transmission to a remote server.
		All of these cards in the chassis can be synchronized using PXI timing and triggering resources. 
		Additionally, a simple DAS software is developed to display the pulsed signals captured and communicate with the host PC for remote control and data upload.
		After careful calibration, preliminary tests show that every DAQ channel achieves an analog bandwidth higher than 200MHz and an ENOB of more than 9 bits at a 1GS/s sampling rate. 
		Owing to such high speed and resolution, the system may facilitate improvements in extracting maximum information from transient signals. Furthermore, with great scalability and high-speed data transmission, the system can be used for other nuclear physics experiments.
	\end{abstract}
	
	
	\IEEEpeerreviewmaketitle
	
	\section{Introduction}
	The topic of ultrafast, high-resolution data acquisition remains one of the most active areas in measurement systems for many modern nuclear physics experiments.
	In this paper, we focus on "the transient pulsed signals produced by detectors in a high-intensity pulsed radiation field"\cite{cheng2010design}. 
	More specifically, this work aims at improving the accuracy of the parameter relevant to the derivation of these signals. 
	Such large dynamic range signals are always one-shot and non-repeatable, and their rising edges can be extremely fast within a range of ten to several nanoseconds, requiring an ultrafast sampling rate of basically at least 1GS/s to digitize and record them. 
	In addition, new challenges to gain more scalability have arisen as channel density is becoming an increasing area of concern due to the huge scale of experiments.
	
    Previous data-acquisition-system (DAS) designs have primarily concentrated on repeatable signals, for which the interested parameters mainly are event counts, energy, signal shape, and arrival time.
    For some particle-physics experiments, the digitization speeds are typically not higher than 100 MS/s \cite{ tang2015upgrade , sbarra2013data}, while in underground neutrino observatory digitization speeds can reach 1GS/s or more \cite{xuetao}.
    However, in the diagnosis of high-intensity pulsed radiation fields, the required speed can be even higher.
    Moreover, in the meantime, our application requires more accuracy than other experiments\cite{cheng2008design}.
    Besides, in order to get the captured waveform data safely, severe destructive electromagnetic and mechanical effects should be paid attention to \cite{cheng2008design,cheng2010design}.

    Conventional ultrafast DASs have been implemented with flash or time-interleaved analog-to-digital convertors (ADCs) with a typical resolution of 8 bits \cite{cheng2010design,liu2014development,zhao20131}, but at the price of cost and power consumption. 
    Recent developments in switched-capacitor-array (SCAs) ASIC chips, achieving sampling rates far beyond 1GS/s at a reasonable cost, have gained more popularity for low-event-rate cases\cite{drs_lecture}. 
    The main drawbacks of SCA ASICs for our application are long dead-time and a lower ENOB parameter\cite{ wang2012waveform, zhao20131}. 
    The application proposed herein focuses on the recording of one-shot fast transient pulses with improved accuracy than the DAS reported in \cite{cheng2010design,cheng2008design,liu2014development}. 

    In this work, we combined a 12bit ADC with a Xilinx FPGA to design a single DAQ card to realize ultrafast data acquisition with improved resolution. Subsequently, we formulated the system to improve its scalability based on the PXI platform with data stored on a local disk via  PXI bus or transmitted to a remote server using the Aurora protocol via SFP+ interface.
	
	\section{Design of Single DAQ Card}

	Considering essential requirements of recording a transient signal, the single DAQ card proposed in this work has the following specifications:
	\begin{enumerate}
		\item Size: 160$\times$100 $mm^2$, conforming with PXI standard regulations
		\item Number of Input Channels: two signal-acquisition channels plus one trigger input
		\item Input Impedance: 50\,$\Omega$ for both signal input and trigger input.
		\item Sampling Rate: 1GS/s
		\item Resolution: 12bit
		\item Record Time: 65\,$\mu s$
		\item Communication Interface: PXI and SFP+
	\end{enumerate}

\begin{figure}[H]
		\centering
		\includegraphics[width=3.5in]{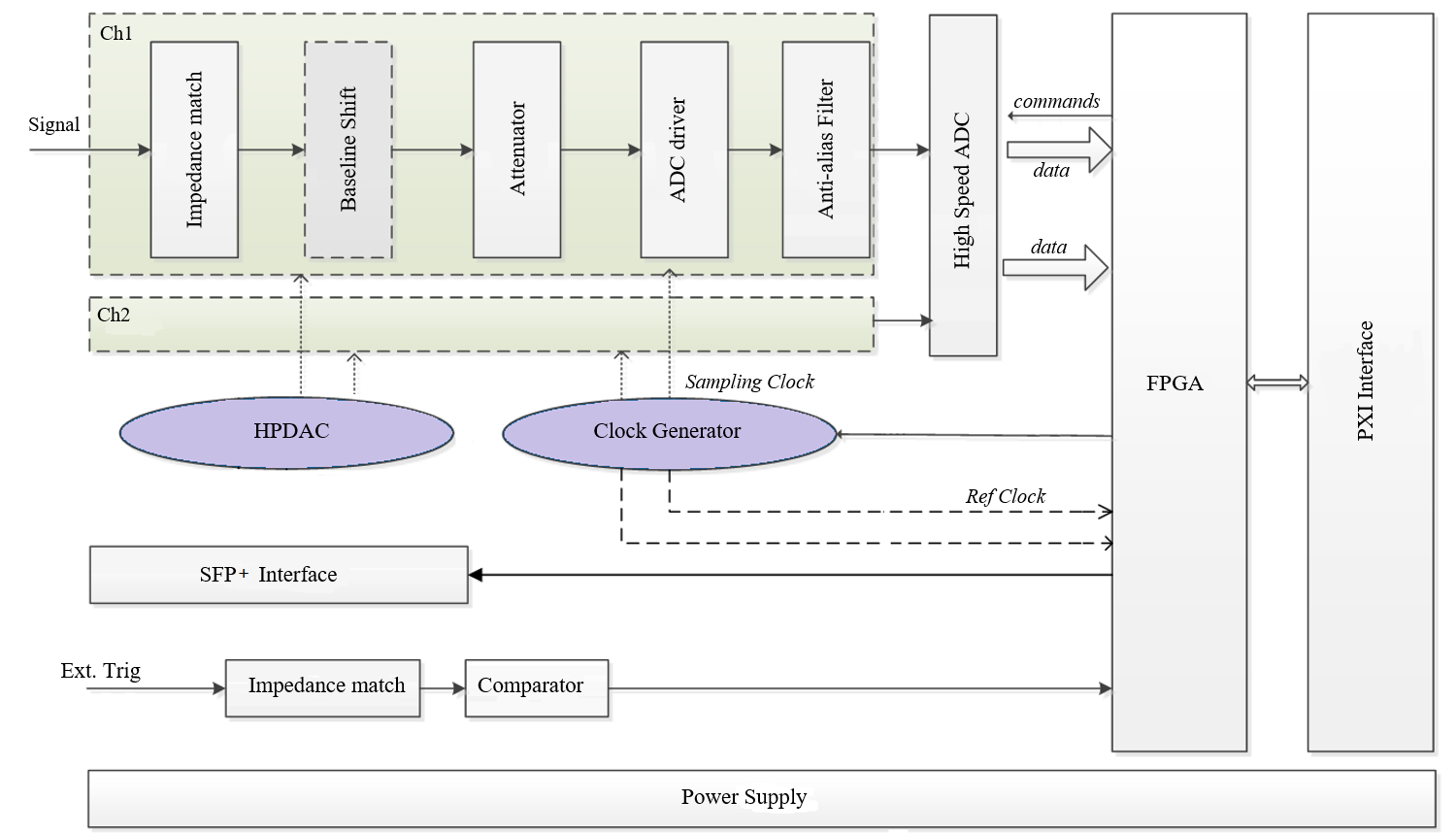}
		\caption{Block diagram of single DAQ card.}
		\label{block_diagram}
	\end{figure}

	The hardware block diagram of a single card is presented in Fig.\ref{block_diagram}, consisting of signal conditioning (also an analog front end), data-acquisition and transmission, power supply, and other peripheral modules.
	
	\subsection{Signal Conditioning Module}
	
	The amplitudes of fast transient pulses can be distributed from some hundreds of milli-voltages to hundreds of voltages. The typical and direct solution is to employ several orders of input range to cover such broad ranges. In the proposed design, a simple '$ \pi $' attenuator network can function as different input-range selection by replacing responding resistors.
	
\begin{figure}[H]
		\centering
		\includegraphics[width=3.3in]{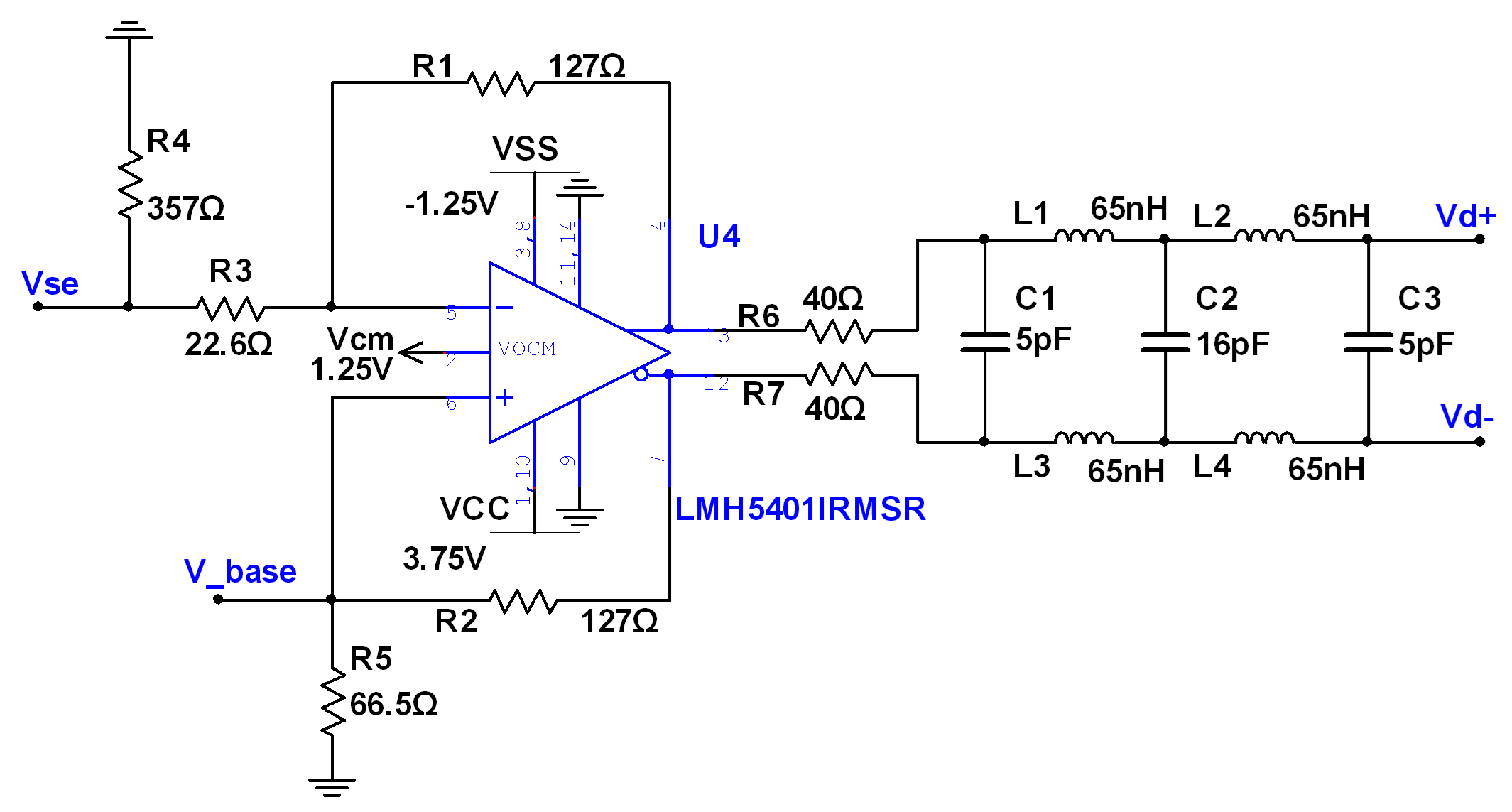}
		\caption{Schematic for LMH5401 implementing signal  SE-DE conversion with a differential Butterworth filter.}
		\label{single_to_diff}
	\end{figure}

	On the differential input stage, We selected a LMH5401 device as the ADC driver to implement single-ended-to-differential (SE-DE) conversion; a schematic is shown in Fig.\ref{single_to_diff}\cite{lmh5401_datasheet}. The accurate value of each corresponding resistor can be calculated according to \cite{adc_driver}. In addition, a five-order differential Butterworth filter\cite{df} was used to act as an anti-alias filter to eliminate degradation of dynamic performance due to high-frequency noises.

\begin{figure}[H]
		\centering
		\includegraphics[width=3.3in]{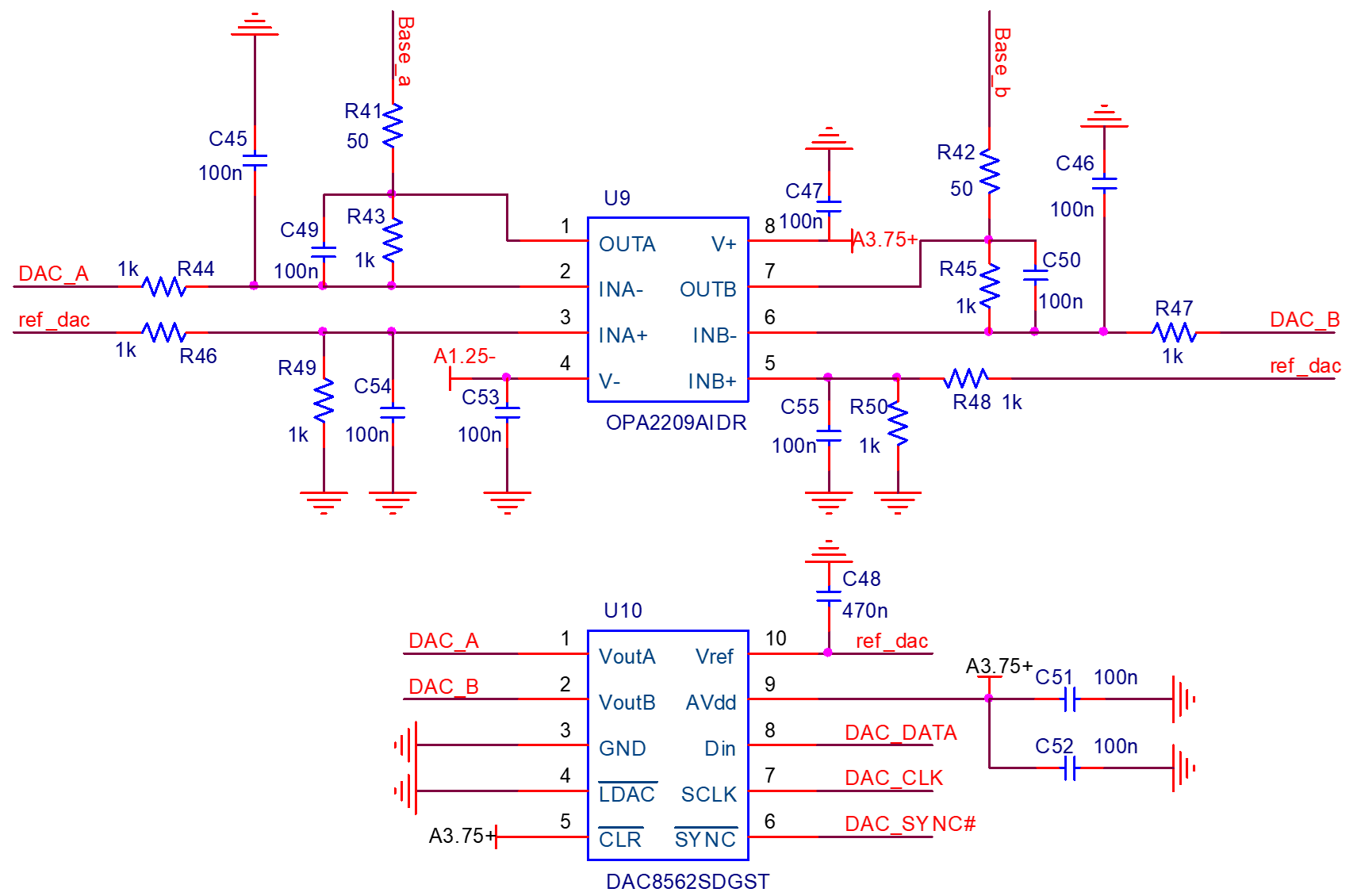}
		\caption{ Schematic for baseline shift module}
		\label{fig:baselineshift}
	\end{figure}

	In our work, we initially designed for unipolar fast transient pulses with possible trailing-edge undershoot or overshoot. However, the ADC chip is inherently a bipolar input stage. This leads to the need for flexible adjustment of the baseline, so we can shift a large positive or negative pulse to the position at which the signal can span roughly the full input range of the ADC. To fulfill the task, a high-precision (16bit) DAC (DAC8562) whose output value is programmed by FPGA was used accompanied by a voltage buffer (OPA2209), and the schematic is illustrated in Fig.\ref{fig:baselineshift}.
	
	\subsection{Data-acquisition and transmission module}
	
	The data-acquisition and transmission module mainly realizes sampling clock generation, analog-to-digital conversion, data processing, and transmission.
	
	The 1 GHz sampling clock demands ultralow root-mean-square (RMS) jitter ($ t_{j} $) to achieve high resolution, as can be seen in (\ref{jitter_SNR}) \cite{kester2010mt008}. We chose a LMK04821 device as the frequency synthesizer chip and a CVHD-950 VCXO as the external reference oscillator to generate this ultralow noise differential (LVDS)  clock. This device integrates two high-performance PLL architectures with three low-noise reference oscillators and consequently generates a very flexible, ultra-low-noise signal at a frequency of, at most 3080MHz \cite{lmk04821_datasheet}.
	
	\begin{equation}
	\label{jitter_SNR}
	SNR_{t_{j}}=20\lg\left[ \frac{1}{2\pi ft_{j}} \right]. 
	\end{equation}
	
	We also selected the ADC12D1000 chip to realize the digitization of dual-channel analog signals at a 1GS/s sampling rate. The ADC device provides a flexible LVDS interface that has multiple SPI programmable options to facilitate board design and FPGA/ASIC data capture \cite{ADC12D1800_datasheet}. We designed the 33MHz/32bit PCI interface for fundamental data communication. Additionally, we employed a SFP+ transceiver (a Finisar FTLF1428P2BNV) to establish a high-speed optical-fiber channel.
	
\begin{figure}[H]
		\centering
		\includegraphics[width=3.4in]{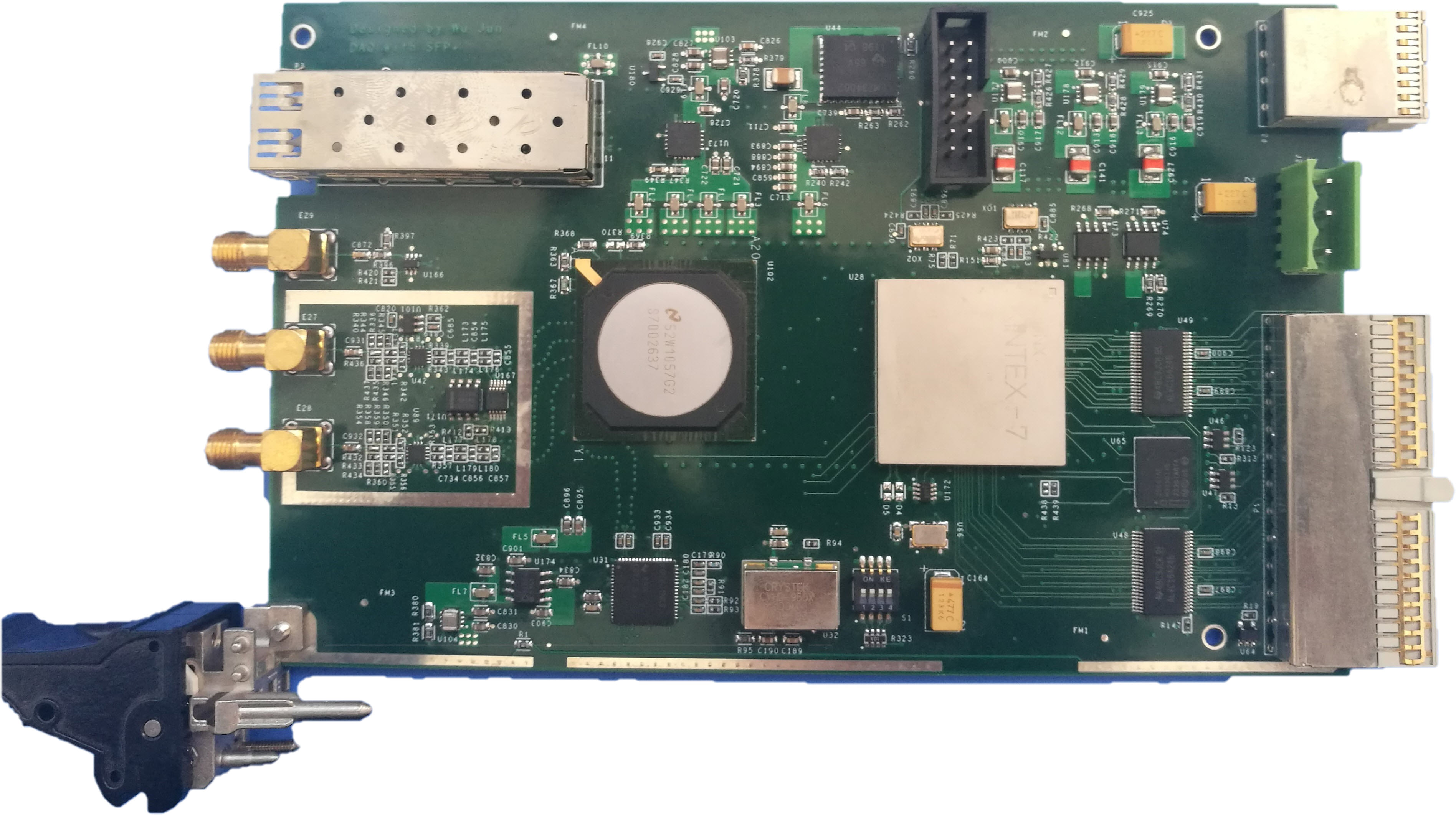}
		\caption{ Photograph of single DAQ card.}
		\label{daq_board}
	\end{figure}

	A photograph of our single DAQ card is presented in Fig.\ref{daq_board}.

	\subsection{Firmware Design with Kintex-7 FPGA}

\begin{figure}[H]
		\centering
		\includegraphics[width=3.5in]{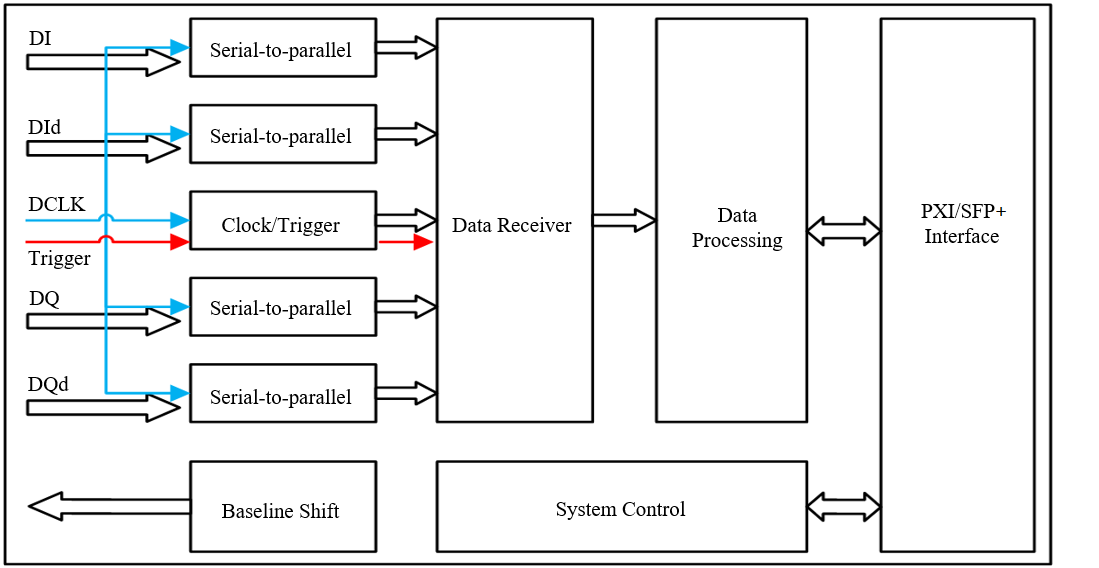}
		\caption{ Basic procedures of firmware design for single DAQ card. }
		\label{procedures}
	\end{figure}

	The designed single DAQ card employed a Kintex-7 FPGA (xc7k160tffg676) for the critical function of capturing the high-speed digital data sourced by the ADC, as well as for configuring other devices with individual interfaces. Basic procedures of firmware design including data receiving and processing are shown in Fig.\ref{procedures}.  
	  
\begin{figure}[H]
		\centering
		\includegraphics[width=3.4in]{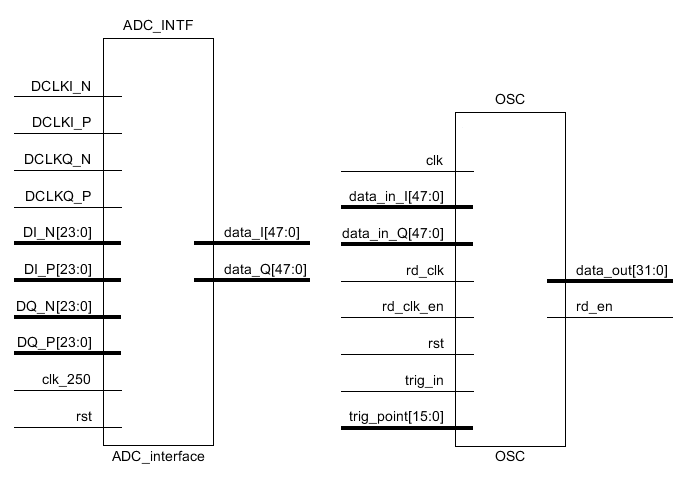}
		\caption{Symbols of ADC interface and data assembly module in firmware design. }
		\label{dam}
	\end{figure}

	Considering ADC data collection and assembly for data transmission,  Fig.\ref{dam} illustrates the ADC interface schematic symbol for high-speed data capture. Instead of dual-port FIFO, we used the integrated Block RAM IP core to instantiate a module to assemble the captured data. Furthermore, data transmission via the PXI interface for local data storage as well as a SFP+ transceiver with optic fibers for further processing could also be designed in the Xilinx Vivado environment using corresponding IP cores. Here, we implemented the Xilinx Aurora 8B/10B IP core using Kintex-7 FPGA GTX transceivers, and thus can realize a data rate of 4.25  Gb/s via a fiber-optic channel.	
	
	\section{System Built on PXI Platform}  
\begin{figure}[H]
		\centering
		\includegraphics[width=2.8in]{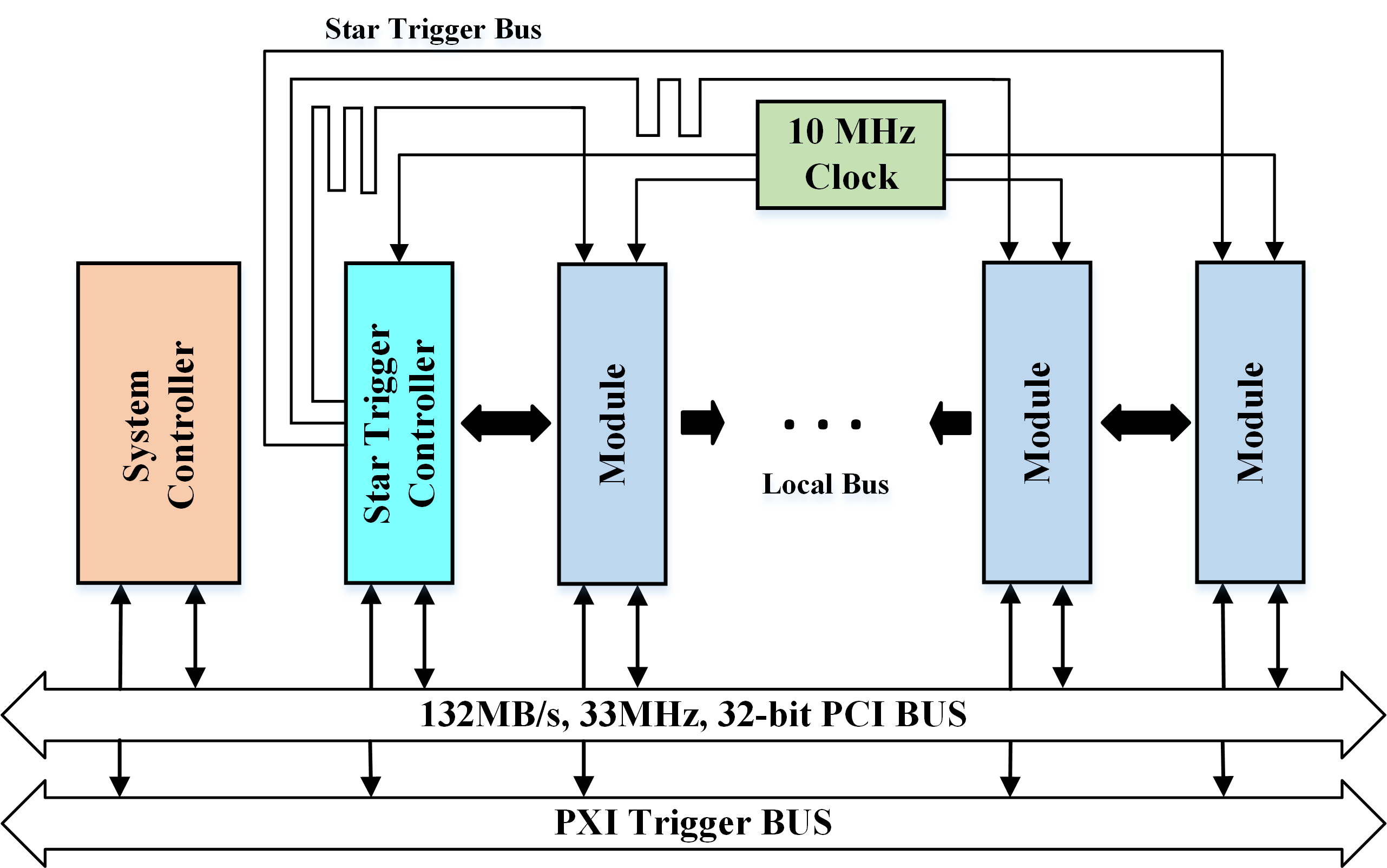}
		\caption{Electrical resources in a PXI 3U chassis.}
		\label{pxi_resource}
	\end{figure}

\begin{figure}[H]
		\centering
		\includegraphics[width=3.4in]{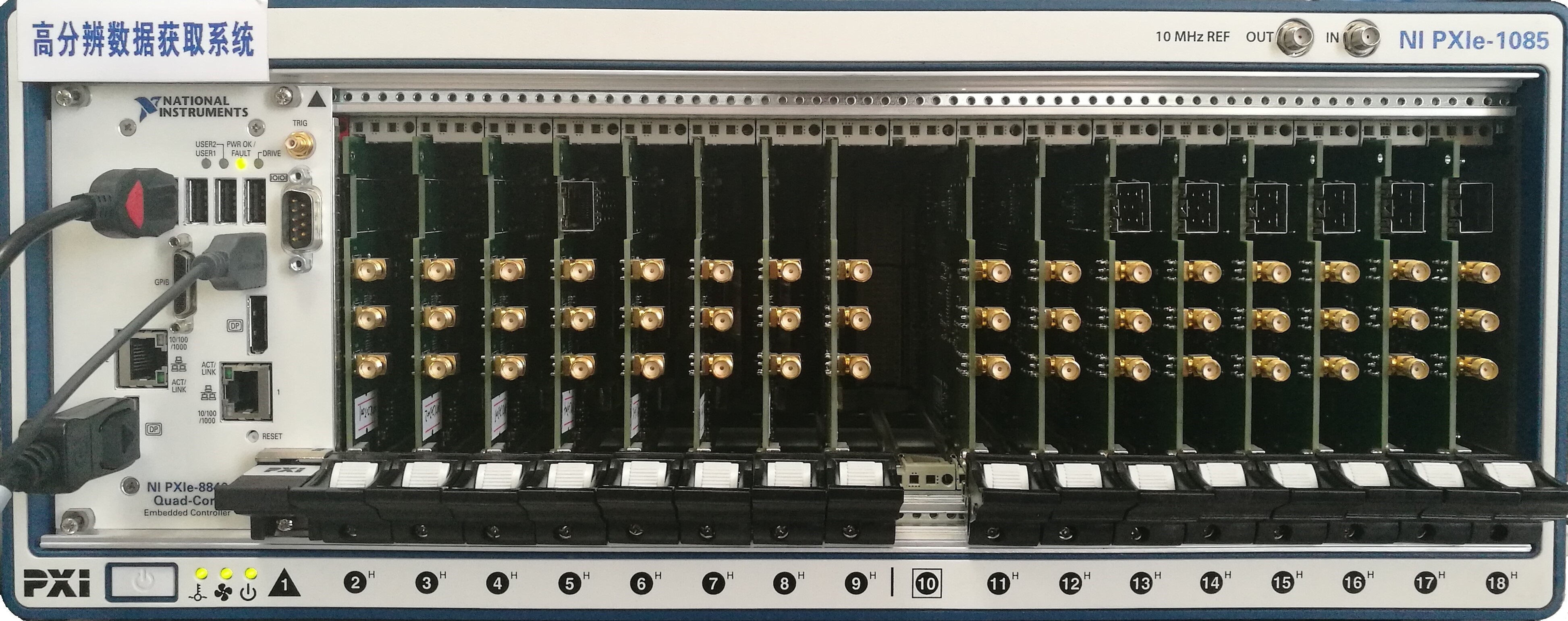}
		\caption{Photograph of PXI-based, Multi-channel, ultrafast DAS for fast transient pulses.}
		\label{DAS}
	\end{figure}

	Every single card in the DAS was designed according to PXI specifications. We built our system from all of these DAQ cards with a NI PXIe-1085 chassis. The 18-slot chassis features an all-hybrid backplane to meet the requirements of 33MHz/32bit PCI bus communication and PXI timing and triggering resources. Fig.\ref{pxi_resource} presents the essential electrical resources on a PXI platform \cite{ni_pxi}. Furthermore, the system can integrate at most 16 DAQ cards in the crate, as presented in Fig.\ref{DAS}. 
	
	The large one-shot fast transient pulse was duplicated and split into 32 signal paths at the experimental site using many external power dividers and onboard '$ \pi $' attenuators. We then connected all 32 channels to the DAS for data digitization and transmission. If the DAQ system is put in the safe area, these data can be stored on the local disk through the PXI bus. But when the system is put near the pulsed radiation source, we must transmit these data through SFP+ optical fiber channels, before the system could be damaged.
	
	The trigger signal comes from other external parts of the experiments and has no direct relationships with our signal to be digitized. Therefore, in this work the hardware part of our DAQ system only implements data transmission; there is no need for data compression due to the event being a one-shot event. In addition, the proposed DAS supports a trigger-position adjustment function similar to that in a commercial oscilloscope through FPGA firmware design.
		
	To realize synchronization of all 16 DAQ cards depicted in Fig.\ref{DAS} through the PXI trigger bus, we configured trigger source and direction in the NI MAX control panel. In addition, we could also realize synchronization through the Star trigger bus in the presence of a dedicated trigger distribution card located in slot 10 of the chassis.

\begin{figure}[H]
	\centering
	\includegraphics[width=3.4in]{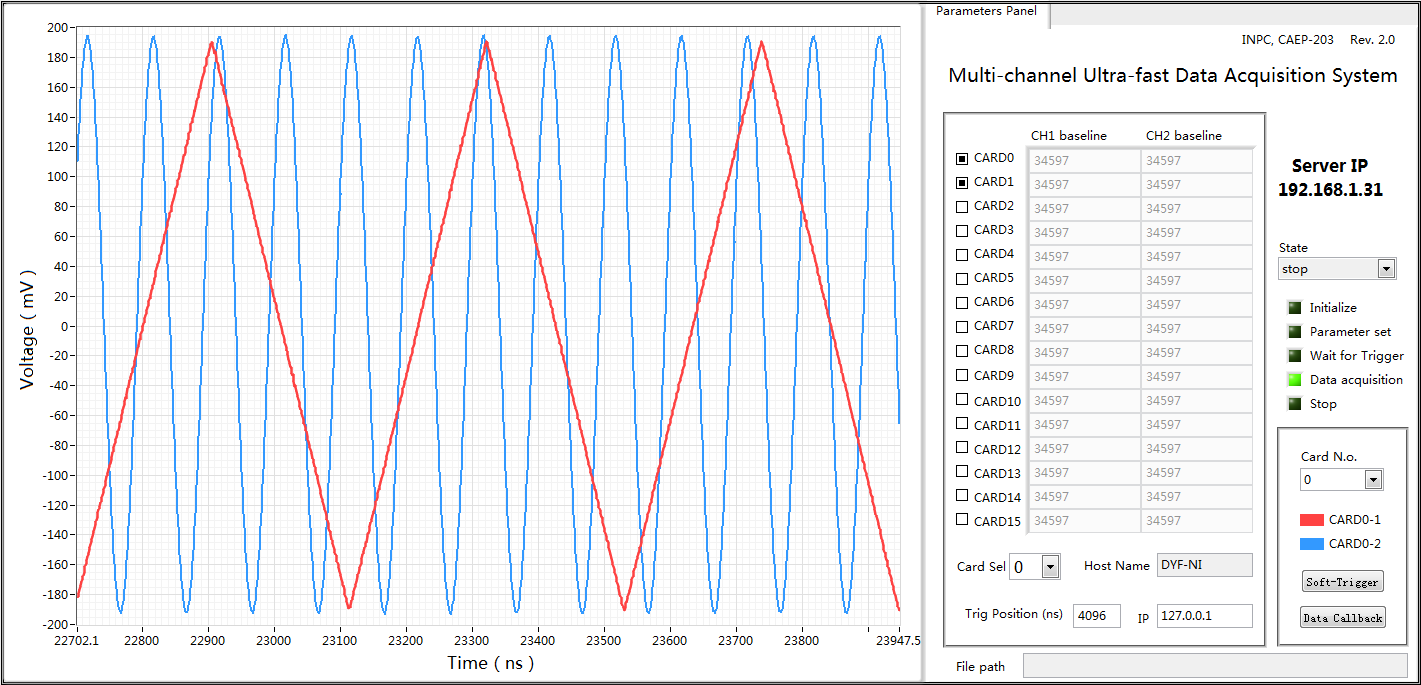}
	\caption{Interface of control software for proposed DAS designed in LabVIEW environment.}
	\label{daq_soft}
    \end{figure}

	NI-VISA provides communication drivers for a variety of programmable instruments, including GPIB, USB, TCP/IP, VXI, RS-232, and PXI instruments\cite{ni_visa}. 
	After configuring VISA in NI LabVIEW environment to create the VISA-based driver and developing basic communication and DMA support\cite{ni_visa_gd}, we built our control software for the DAS as depicted in Fig.\ref{daq_soft}. With this software, we can implement controls of baseline adjustment, trigger-position set, and waveform display. Furthermore, the software can also function as an interface to the remote server.  
	
	\section{Preliminary Tests and Results}	
	
\begin{figure}[H]
		\centering
		\includegraphics[width=3.6in]{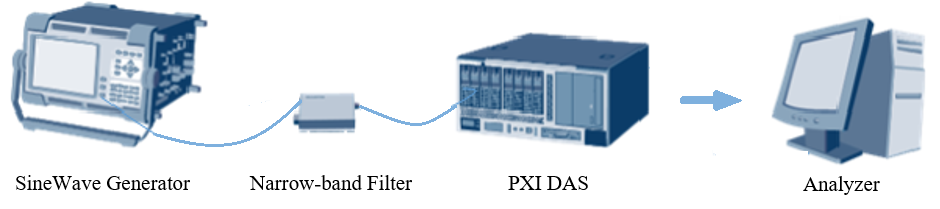}
		\caption{DAS test architecture.}
		\label{test_archi}
	\end{figure}

	Typically, the full-scale-range (FSR) input range of the ADC is 800 mV (differential mode voltage) by default and can be 1 V through register configuration.
	We configured the '$ \pi $' attenuator network and the amplifier by using corresponding resistors to set FSR of a single channel to be 400 mV, and then implemented several key tests. Typical test setups are shown in Fig.\ref{test_archi}. Here, the LC band filters were acquired from a local microwave scientific company (http://www.blmicrowave.com.cn). 
	
	\subsection{BandWidth Test}
	
\begin{figure}[H]
		\centering
		\includegraphics[width=3.3in]{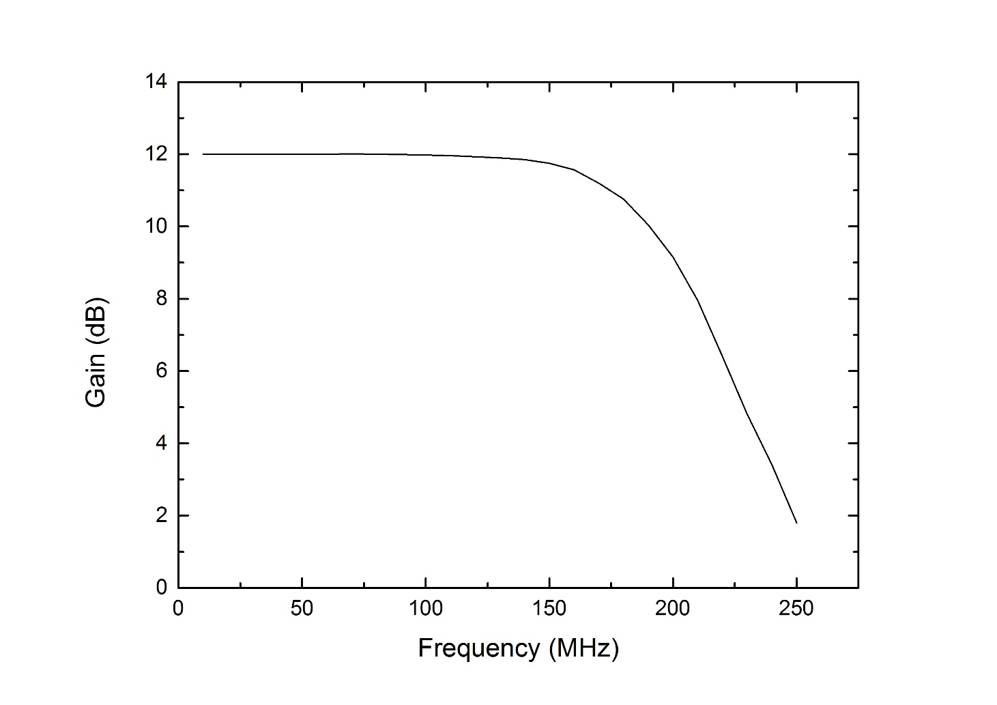}
		\caption{Typical gain-vs-f plot of a single channel.}
		\label{bandwidth}
	\end{figure}

	First, we tested the bandwidth of each single DAQ card. Considering that the -3dB bandwidth of the Butterworth filter is higher than 200MHz, typical test result shown in Fig.\ref{bandwidth} is consistent with this point. In addition, the bandwidth for -0.1dB flatness achieved a value of roughly 130MHz.
	
	\subsection{ENOB Test}
	ENOB is mostly used to evaluate the overall dynamic performance of an ADC or ADC-based measurement system. According to the IEEE-1241 and IEEE-1057 standards, we implemented preliminary tests for ENOB at four frequency points. Typical results can be seen in Tab.\ref{enob}, with the last column listing the respective SINAD parameters. Additionally, the RMS aperture jitter can also be estimated from (\ref{jitter_SNR}).
	
	\begin{table}[H]
		\renewcommand{\arraystretch}{1.3}
		\caption{Typical ENOB test results at four frequency points.}
		\label{enob}
		\centering
		\begin{tabular}{c|c|c}
			\hline
			\bfseries Frequency (MHz)& \bfseries ENOB (bit) & \bfseries SINAD (dB) \\
			\hline
			23.723   & 9.3   & 57.75 \\
			58.541   & 9.3   & 57.75 \\
			134.337  & 9.2   & 57.14 \\
			190.431  & 8.6   & 53.53 \\
			\hline
		\end{tabular}
	\end{table}
	
%

	\subsection{Recording of Fast Transient Pulses}
\begin{figure}[H]
		\centering
		\includegraphics[width=3.3in]{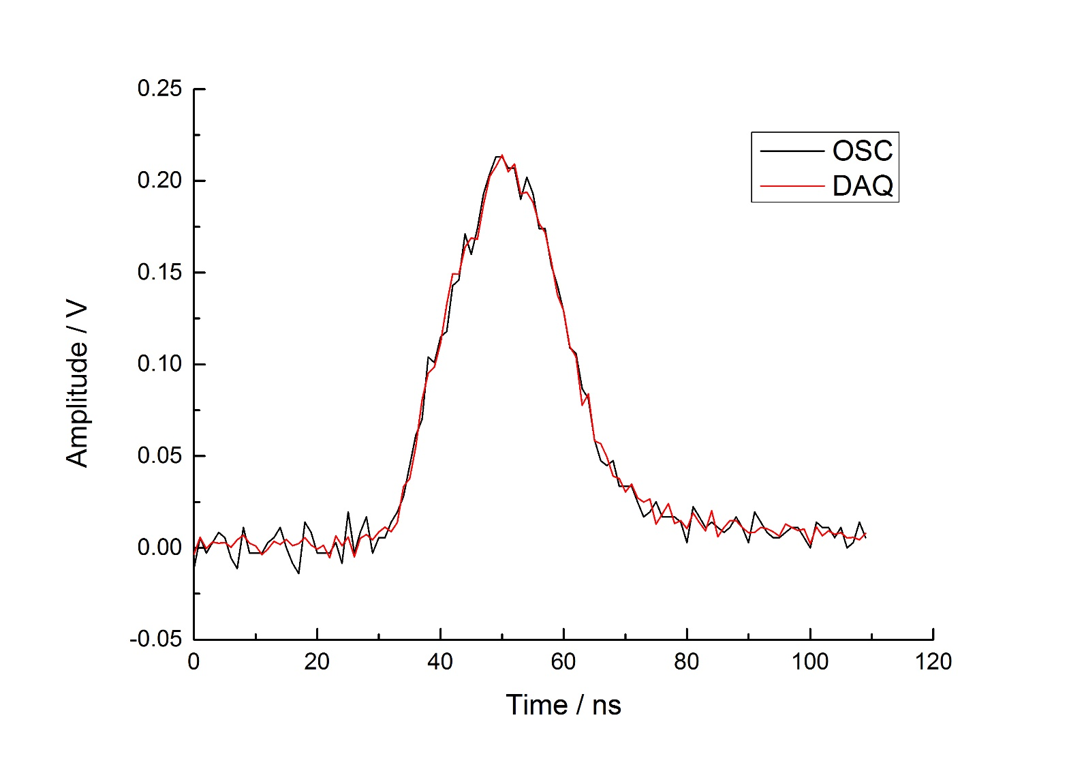}
		\caption{Typical comparison between proposed DAQ channel and a Lecroy HDO6054 oscilloscope channel when recording a typical one-shot fast transient pulse.}
		\label{pulse_record}
	\end{figure}

	To test the performance difference on recording fast transient pulses between the proposed DAS and a Lecroy HDO6054 oscilloscope, we conducted relevant experiments at China's Northwest Institute of Nuclear Technology (NINT), and the results are shown in Fig.\ref{pulse_record}. The comparison of these two plots clearly shows that the waveform captured by the proposed DAS is much smoother. This means that we can achieve better uncertainty when fitting to these data to obtain a mathematical formula. The difference revealed by experiment demonstrates that the proposed DAS has improved accuracy in recording fast transient pulses compared to a commercial oscilloscope in this test.
	
	In addition, we successfully realized baseline adjustment through software and transmitted the digitized data to the remote server within 1 ms through the SFP+ interface at a speed of 4.25  Gb/s (in this test, another of the same DAQ card acted as a receiver).
	Owing to such high data-transmission speed, the sampling data can be transmitted to the remote server before the system could be damaged by the severe destructive electromagnetic and mechanical effects in microseconds\cite{cheng2008design,cheng2010design}.
	
	\section{Conclusions and Outlook}
    The proposed DAS can integrate 32 channels into one chassis. Preliminary measurements show that a single channel in the proposed DAS can achieve a -3-dB bandwidth higher than 200MHz and an ENOB of more than 9 bits at a 1GS/s sampling rate. Each channel has a memory depth of 65 kS, and the trigger position can be programmed through the software interface. The digitized data can be either transferred through the PXI 33MHz/32bit bus to the disk of a local controller or through the SFP+ interface to the remote receiver safely.
    Besides, features of input range selection and baseline adjustment were designed to satisfy the variable and unpredictable amplitude and unipolarity of the signal. 
    All of these points contributed to the improved adaptability in our applications and other nuclear physics experiments.

   This work was the preliminary efforts to invent ultrafast DASs with improved accuracy for one-shot fast transient pulses. In the recent future, increasingly improvements on more accurate, greater scalability, higher sampling and transmission speed are planned. Moreover, instruments or setups are typically designed according to test theories and methods based on periodic signals that are probably not a likely scenario in the field of measuring one-shot transient pulsed signals. Therefore, further relevant research work will also be done in the future.

\end{document}